\documentclass[sigconf,authorversion,nonacm]{acmart}

\AtBeginDocument{%
  \providecommand\BibTeX{{%
    \normalfont B\kern-0.5em{\scshape i\kern-0.25em b}\kern-0.8em\TeX}}}

\usepackage{xcolor,pifont}
\definecolor{cvprblue}{rgb}{0.21,0.49,0.74}
\usepackage{hyperref}
\hypersetup{
  colorlinks   = true, 
  urlcolor     = blue, 
  linkcolor    = red, 
  citecolor = cvprblue
}

\usepackage{multirow}

\graphicspath{{./images/}}

\begin{document}

\title{Who Checks the Checkers?\\ Exploring Source Credibility in Twitter's Community Notes}

\author{Uku Kangur}
\orcid{0009-0005-9393-6848}
\affiliation{%
  \institution{University of Tartu Institute of Computer Science}
  \city{Tartu}
  \country{Estonia}}
\email{uku.kangur@ut.ee}

\author{Roshni Chakraborty}
\affiliation{%
  \institution{University of Tartu Institute of Computer Science}
  \city{Tartu}
  \country{Estonia}}
\email{roshni.chakraborty@ut.ee}

\author{Rajesh Sharma}
\affiliation{%
  \institution{University of Tartu Institute of Computer Science}
  \city{Tartu}
  \country{Estonia}}
  \email{rajesh.sharma@ut.ee}



\begin{abstract}

In recent years, the proliferation of misinformation on social media platforms has become a significant concern. Initially designed for sharing information and fostering social connections, platforms like Twitter (now rebranded as X) have also unfortunately become conduits for spreading misinformation. To mitigate this, these platforms have implemented various mechanisms, including the recent suggestion to use crowd-sourced non-expert fact-checkers to enhance the scalability and efficiency of content vetting. An example of this is the introduction of Community Notes on Twitter. 

While previous research has extensively explored various aspects of Twitter tweets, such as information diffusion, sentiment analytics and opinion summarization, there has been a limited focus on the specific feature of Twitter Community Notes, despite its potential role in crowd-sourced fact-checking. Prior research on Twitter Community Notes has involved empirical analysis of the feature's dataset and comparative studies that also include other methods like expert fact-checking. Distinguishing itself from prior works, our study covers a multi-faceted analysis of sources and audience perception within Community Notes. We find that the majority of cited sources are news outlets that are left-leaning and are of high factuality, pointing to a potential bias in the platform's community fact-checking. Left biased and low factuality sources validate tweets more, while Center sources are used more often to refute tweet content. Additionally, source factuality significantly influences public agreement and helpfulness of the notes, highlighting the effectiveness of the Community Notes Ranking algorithm. These findings showcase the impact and biases inherent in community-based fact-checking initiatives. 

\end{abstract}





\maketitle
\section{Introduction}
\label{sec:intro}
Misinformation on social media platforms has become a pressing issue which has captured 
the attention of Governments, policymakers, researchers and the general public \cite{Rainie2017}. Several studies indicate that false information spreads faster than verified facts. This is particularly true for Twitter where information irrespective of its truthfulness can spread across a huge fraction of the population in a small duration of time \cite{vosoughi2018,chakraborty2017network}. This rapid dissemination of false information poses a serious concern as it has tangible and impactful real-world consequences, such as, on public health, elections, and national security. For example, one of the most well-known instances that illustrates the severity of the issue in recent times was a series of tweets from former U.S. President Donald Trump where he falsely claimed that the 2020 US presidential elections were faked \cite{Yen2020}. This misinformation played a significant role in inciting a violent attack on the U.S. Capitol on January 6, 2021 \cite{BBC2021}. 




Misinformation detection is the foremost step in mitigation of misinformation. Traditional methods often involve expert fact-checkers or specialized organizations that use their expertise to validate or debunk claims made on digital platforms \cite{MorenoGil2022}. However, this approach faces scalability issues as the volume of online content far exceeds the capacity of such experts to scrutinize it \cite{marcos2021}. An alternative strategy involves the use of automated misinformation detectors \cite{sharma2021identifying},
such as, machine learning techniques \cite{guo-etal-2022-survey} which analyze the underlying linguistic patterns to distinguish between truthful and misleading content \cite{butt2022goes}, \cite{sharma2022mis}. Despite their utility, these automated systems are not infallible and frequently necessitate human moderation for optimal performance \cite{Horne2023} and further, provide suggestions to combat misinformation \cite{shakshi2023mis}, \cite{sharma2023amir}.


To tackle both of these challenges, social media platforms have moved more towards reliance on the wisdom of crowds for fact-checking \cite{allen21}. In 2021, Twitter introduced a community-based fact-checking service originally known as Birdwatch, which was subsequently rebranded as Community Notes \cite{biron2022musk}. While the service was initially built to add useful context to tweets, it has recently taken a shift towards combating misinformation on the platform \cite{duffy2022dorsey}. Community Notes allows users to flag misleading tweets and provide annotations that include context or corrections. Users have the option to further substantiate their notes by adding links to external sources. Experts on misinformation believe that partisanship is the main reason for spreading misinformation \cite{altay2023}. Due to this, it is crucial for social meia platforms to minimize bias, especially in the context of fact-checking. 

\begin{figure}[t]
\centering
\includegraphics[width=1\columnwidth]{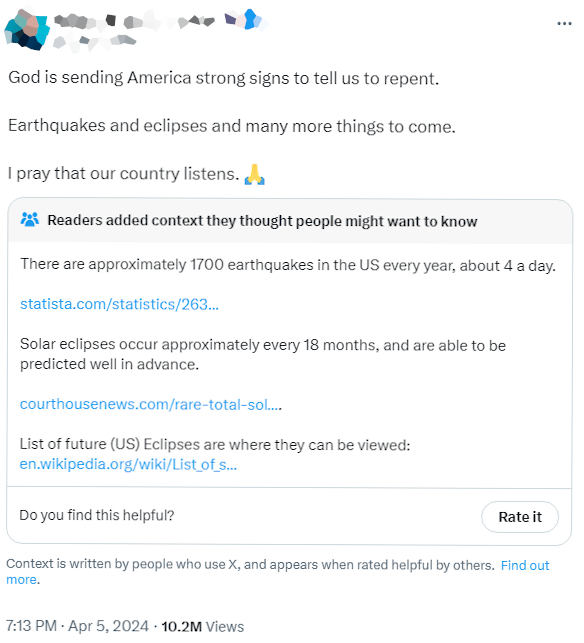}
\caption{Screenshot of a tweet and its associated Community Note. The note itself cites sources used to fact-check the claim.}
\label{comm-note}
\end{figure}

The aim of this research is to shed light on the community-driven aspect of fact-checking within Twitter's Community Notes feature. With the increase in prevalence of misinformation, particularly in politically charged arenas, understanding the dynamics of community-based fact-checking is more critical than ever. While a few works have looked at community note user interactions and user consensus \cite{Prollochs_2022, pilarski2023community}, none have yet explored the sources used in Community Notes (See Figure \ref{comm-note}). This research benefits platform developers, policymakers, and researchers by providing actionable insights into the sources and potential biases that may skew public discourse. We systematically evaluate the types of sources cited, analyze their biases, and factuality levels, and probe the impact of these variables on audience perceptions. 
We summarize the contributions of the study through the following research questions (RQs): 

\noindent\textbf{RQ1 - Sources of Validity:} \textit{How do the citation patterns within Community Notes reflect and influence the biases and factual integrity of the information that shapes public opinion and fact-checking efforts?} 
With this research question, we examine the patterns of web page citations within Community Notes to uncover how information sources influence public opinion and fact-checking efforts. Initially, we identify the most frequently cited web pages and investigate whether citation frequencies differ based on the country of origin of these pages, aiming to understand geographical biases in source popularity. Subsequently, we evaluate the bias and factuality levels of these commonly cited web pages to assess their impact on shaping collective viewpoints. This evaluation is crucial for identifying potential misinformation and understanding the overall reliability of the information disseminated. Furthermore, we explore correlations between the type of source, its bias, and its factuality, seeking to reveal systematic tendencies in the selection and use of sources. By examining these aspects, we aim to highlight areas where critical evaluation of sources is needed to enhance the credibility and factual grounding of shared information.
    
\noindent\textbf{RQ2 - Perceptions of audience:} \textit{How do source characteristics such as type, bias, and factuality impact their effectiveness in refuting or supporting Community Notes and influence the perceived helpfulness and agreement with them?} 
In this research question, we investigate the roles that different categories of sources play in supporting or refuting content shared on social media platforms, specifically Twitter. Furthermore, we examine how these same attributes, source type, bias, and factuality, affect the perceived helpfulness of explanatory notes appended to tweets. This exploration seeks to determine whether notes rated as helpful are also those that are unbiased and fact-based. Additionally, we study the impact of source characteristics on the perceived agreement with the notes, assessing whether source credibility influences audience perceptions. Through this comprehensive analysis, we aim to highlight the complex interplay between source credibility and the reception of crowd-based fact-checking on Twitter.
These questions expand our understanding of how community-based fact-checking functions and provide critical insights for platform developers, policymakers, and researchers aiming to improve the efficacy and fairness of online information verification systems. The rest of the paper is structured as follows: In Section \ref{sec:related}, we present an overview of the related works in the domain. In Section \ref{sec:data}, we present the
details about the data used for the experiments. In Section \ref{sec:analysis}, we
introduce the details of the experiments carried out and discuss the re-
sults obtained. The conclusions and Ethics Statement are presented in Sections \ref{sec:conclusion} and \ref{sec:ethics} respectively. We additionally release the code\footnote{The anonymized code of the analysis is available here: \url{https://anonymous.4open.science/r/CN-528C/}} used for the analysis of our results.


\section{Related work}
\label{sec:related}
This section covers past works that deal with exploring bias in fact-checking and sources of information on social media, which are the focus areas of this study.

\noindent{\textbf{Fact-checking on social media:}} Fact-checking on social media primarily falls into three categories: expert, automated, and community-based \cite{bozarth2022pitfalls}. These branches have evolved to meet the unique challenges posed by the rapid dissemination of information on social media platforms\cite{chakraborty2019tweet,chakraborty2023twminer}. Expert fact-checking platforms, such as PolitiFact \cite{li2023combating} and Snopes \cite{Hannak_Margolin_Keegan_Weber_2014} provide a thorough understanding of news articles and statements as they rely on professional fact-checkers. However, expert fact-checking requires extensive time and effort, which makes it difficult to handle the huge volume of information disseminated online \cite{marcos2021}. Recently, automated fact-checking-based approaches have become popular, which leverage machine learning and natural language processing techniques for instant verification \cite{guo-etal-2022-survey}. Although these methods are highly scalable, they fail to provide justification and interpret the contextual information, which is essential for reliable and trustworthy fact-checking \cite{Santos2023}, \cite{nikopensius2023reinforcement}. To overcome these challenges, several existing research studies have proposed community-based fact-checking as a possible alternative. Community-based fact-checking is not dependent on a few expert individuals as it leverages the wisdom, allowing for several factual interpretations \cite{Godel_Sanderson_Aslett_Nagler_Bonneau_Persily_A.Tucker_2021}. However, while it solves the issues of fact-checking speed and explanability, it can still be susceptible to biases and manipulation \cite{Saeed_2022}.


\noindent{\textbf{Bias in Fact-Checking:}} Bias can significantly influence human perception and the creation of fact-checks, particularly on issues that evoke strong negative opinions \cite{park2021presence}. Understanding this is crucial, as these inherent biases affect users' interpretation of information. Although several works indicate strong bias in fact-checking users for social media platforms, such as Twitter \cite{shin2017}, this has been explored only on a few expert-based fact-checking platforms, such as PolitiFact \cite{draws2022}. We indicate that no research has studied and analyzed the bias in Twitter community notes, which is one of the reasons this study focuses on the issue.



\noindent{\textbf{Media source bias and factuality:}} Research on media sources is extensive and diverse, with a focus on various aspects that contribute to media bias. One avenue has been to investigate quoting patterns to discern biases \cite{niculae2015quotus}. Another line of research has examined how media bias is evident in the citations to think tanks and policy groups \cite{groseclose05}. More recently, the influence of bias in headlines has also been studied \cite{pan2023bias}. These traditional media studies lay the groundwork for understanding bias, but the landscape is continuously evolving with the growth of social media platforms. On Twitter, scholars have taken different approaches to analyzing media bias. One such method has been to align co-subscribers of news sources to deduce potential biases \cite{an_2021}. Another approach has looked at the nature of reactions in Twitter comments to different news articles, offering insights into public perception and inherent biases \cite{SPINDE2023100264}. Beyond user behaviour, it's also crucial to consider the role of algorithms; for instance, studies have explored how Twitter's algorithm amplifies content with varying degrees of bias \cite{huszar2022}. Given the intricate interplay of user behaviour and algorithmic influence in shaping and amplifying bias, our study aims to delve deeper into the dynamics of community-based fact-checking. 

\noindent{\textbf{Twitter Community Notes:}} Although there are a plethora of research works on Twitter datasets covering topics such as information diffusion \cite{vosoughi2018}, sentiment analysis \cite{wang2022}, content sources \cite{singh2020understanding}, etc., there are very few existing research works on Twitter Community Notes \cite{wojcik2022birdwatch}. Existing works on Twitter Community Notes includes that of Pröllochs et al. \cite{Prollochs_2022}, in which they empirically analyze the Twitter Community Notes by examining user interactions, note credibility, sentiment, and the influence of tweet authors on user consensus. Subsequent research has expanded to include comparative studies that examine Twitter Community Notes with respect to fact-checking, such as snoping and expert reviews \cite{pilarski2023community, drolsbach2023diffusion, Saeed_2022}. However, none of these approaches performs a study of the bias and fact-checking of the sources. Therefore, in this paper, we focus on a multi-faceted exploration of sources and audience perception within Twitter Community Notes. Unlike prior studies that focused on the diffusion, consensus, and sentiment of users in reaction to Community Notes compared to expert fact-checks, our research aims to understand how source characteristics influence audience perceptions, thereby providing a more comprehensive understanding of source credibility's role in shaping public opinion. We observe that \url{mediabiasfactcheck.com}, \url{allsides.com}, \url{adfontes.com}  have been highly effective and reliable in the detection of source bias in both Twitter \cite{Samory_Kesiz_Abnousi_Mitra_2020, huszar2022} and Reddit \cite{Weld2021PoliticalBA, bayiz2024}. Therefore, we utilize these media bias ranking platforms, to identify, understand and interpret bias in Twitter Community Notes which will inherently provide a more nuanced understanding of online crowd-based fact-checking.
\section{Data}
\label{sec:data}

In this section we explain the data used for the study. We additionally highlight the key variables we use to refer to certain features of our data. We cover data related ethical questions under Section \ref{sec:ethics}.

\subsection{Community Notes}

\begin{figure*}[t]
\centering
\includegraphics[width=1\textwidth]{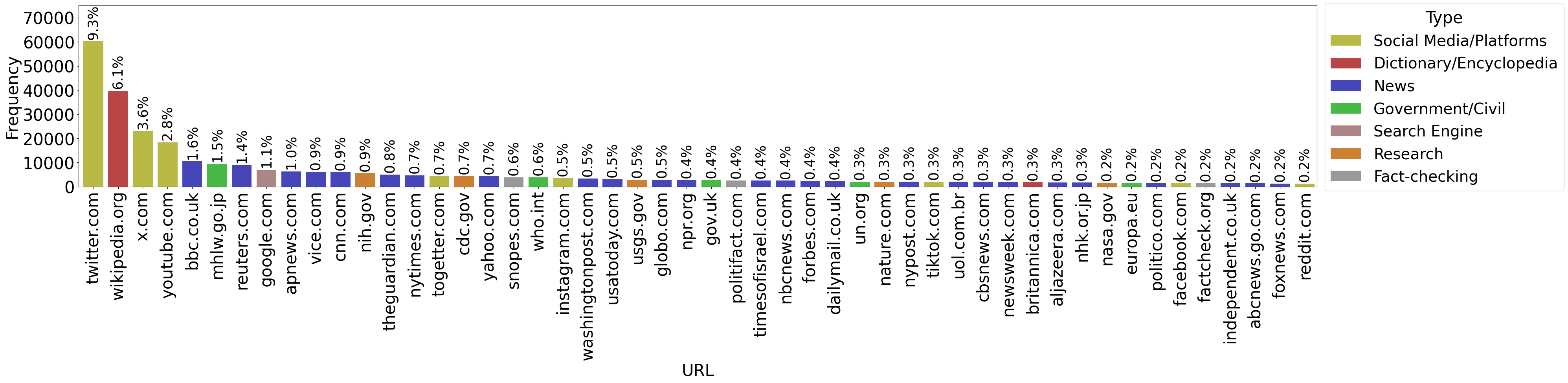}
\caption{Types of top 50 sources (after combining related domains). The relative frequencies over the whole dataset are plotted on top of each URL bar.
}
\label{top-cat}
\end{figure*}

Community Notes (previously called Birdwatch) is a community-driven fact-checking feature on Twitter which was launched in January $2021$ \cite{xcommunitynotes}. The fact-checks on Community Notes are called "notes" in short. To write notes, Twitter users must separately sign up for the Community Notes feature. Once verified, Community Note users can start reviewing tweets, writing notes and rating other Community Note users' notes. For the Twitter end-user, only the highest-rated Community Note per tweet is displayed (the Community Note must also have at least $5$ ratings to be displayed). In addition, Twitter end-users can rate the top displayed note of a tweet. The minimum rating threshold restricts the spreading of bot-generated notes. 

\noindent{\textbf{Data Collection:}} We downloaded all of the publicly available (royalty-free) Community Note data from the Twitter Community Note website\footnote{The data is available here: \url{https://communitynotes.twitter.com/guide/en/under-the-hood/download-data}} \cite{twitter_2023} from the period starting from January $23$, $2021$ and ending with January $27$, $2024$. The dataset is separated into four subsets. The first subset, \textbf{Notes}, contains information about all notes. The second subset, \textbf{Ratings}, contains information about the ratings of a note. The third subset, \textbf{Note status history}, contains metadata about notes, including what statuses they received and when. The final and fourth subset, \textbf{User status}, contains metadata about each user’s enrollment state. 
For our analysis, we used the Notes, Ratings and Note status history sub-datasets. The Notes and Note status history datasets comprise of $544995$ Community Notes from $87294$ users, and the Ratings dataset comprises of $6514542$ Ratings.



\subsection{Key variables}

We introduce the key variables we extracted using data mining from the dataset or additionally annotated.
\textbf{Content} refers to the textual content of the note. This includes any supporting source links added by the Community Notes user as well. \textbf{Source} refers to the hostname of the URL that the author of the note refers to in the corresponding note as the source. \textbf{Type} refers to the type of the source. \textbf{Bias} refers to the bias rating of the source. \textbf{Factuality} refers to the factuality rating of the source.
\section{Analysis}
\label{sec:analysis}

\subsection{Sources of validity (RQ1)}
In this subsection, we employ data collection, cleaning, and categorization techniques to understand the variety of sources cited in Community Notes. Our objective is to identify which publications and their origins are most commonly utilized and assess their contribution to the platform's information veracity through analysing $Bias$ and $Factuality$.

\subsubsection{Cited sources:} 
We, initially, investigate which sources are cited more (compared to others) and analyze the types of these sources. For this, we collect all the web links (44523 unique links in total) from the \textbf{Content} of the notes and then, perform the following pre processing. We initially simplify the URLs to their hostnames, such as reducing \textit{https://www.example.com/article/123} to \textit{example.com}, address issues with short links, redirects, automatically retrieve the original sources from web archive links by using a script and accept redirect links only when originals were unavailable. The complete list of expanded URLs can be seen in Table \ref{tab:categorized_urls}.

\begin{table}[h!]
\fontsize{8pt}{10pt}\selectfont
\begin{tabular}{ll}
\toprule
\textbf{Category} & \textbf{URLs} \\
\midrule
Link shorteners & \url{tinyurl.com}, \url{www.shorturl.at},\\
& \url{bit.ly}, \url{is.gd} \\
Social media redirects & \url{g.co}, \url{t.co}, \url{yahoo.com}, \url{goo.gl},\\ & \url{youtu.be}, \url{redd.it}, \url{fb.me} \\
\textit{Web archives} & \url{web.archive.org}, \url{archive.ph},\\
& \url{archive.is}, \url{archive.org} \\
\bottomrule
\end{tabular}
\caption{List of URLs Expanded}
\label{tab:categorized_urls}
\end{table}

\begin{table*}[t]
\centering
\caption{Most Frequent Sources by Type in Top 500 Sources, including their contribution percentages in their corresponding type. Note that there were only two sources for the Web Archive category as most of them were already expanded.}
\label{tab:top_sources}
\fontsize{8pt}{10pt}\selectfont
\begin{tabular}{lll}
\toprule
\textbf{Type} & \textbf{Top Sources} & \textbf{Percentage of Type} \\
\midrule
\textit{News} & BBC, Reuters, AP & 6.29\%, 5.34\%, 3.81\% \\
\textit{Fact-checking} & Snopes, Politifact, FactCheck & 31.72\%, 21.42\%, 12.13\% \\
\textit{Dictionary/Encyclopedia} & Wikipedia, Britannica, Merriam-Webster & 87.78\%, 4.11\%, 2.36\% \\
\textit{Government/Civil} & MHLW, WHO, Gov.uk & 16.96\%, 7.02\%, 4.92\% \\
\textit{Social Media/Platforms} & Twitter, X, YouTube & 48.54\%, 18.60\%, 14.82\% \\
\textit{Research} & NIH, CDC, USGS & 15.35\%, 12.07\%, 8.08\% \\
\textit{Web Archive} & Wayback Machine, DOI & 50.61\%, 49.39\% \\
\textit{Search Engine} & Google, Justia, Bible Gateway & 92.83\%, 3.13\%, 2.20\% \\
\textit{Other} & all-senmonka.jp, ne.jp, apple.com & 15.39\%, 12.77\%, 8.69\% \\
\bottomrule
\end{tabular}
\end{table*}

Furthermore, after the initial preprocessing, we group similar URLs that represent the same domain. This grouping includes different device versions of websites, such as \url{en.wikipedia.org} and \url{en.m.wikipedia.org}, and both short and long forms of websites like \url{youtube.com} and \url{youtu.be}. Additionally, we consolidate different country versions of websites, exemplified by \url{bbc.com} and \url{bbc.co.uk}. We also group subpages of the same institutional websites, such as \url{twitter.com} and \url{help.twitter.com}, recognizing them as originating from the same core domain. This approach helps our analysis by reducing redundancy and focusing on fundamental source identities. We consider the top $500$ URL groups on the basis of highest occurence frequency in the Community Notes data. These groups comprise of total $4064$ URLs and have been used in $306578$ notes  (approximately 56\% out of all notes in that period). For the rest of the paper, we refer to these groups as top $500$ sources.


We consulted \url{mediabiasfactcheck.com}
supplemented by the sources "About" sections to categorize the top $500$ sources into $8$ \textbf{Type} categories: \textit{News}, \textit{Fact-Checking}, \textit{Dictionary/Encyclopedia}, \textit{Government/Civil}, \textit{Social Media/Platforms}, \textit{Research}, \textit{Search Engine}, \textit{Web Archive} and \textit{Other}. \textit{Other} category comprises of URLs that do not fit into any of the other categories, for example, private business pages, portfolios, download links, etc.
On evaluation of the distribution of source categories, we observe a long-tail pattern, where a small number of sources are extremely frequent while the majority are cited less often as can be seen on the percentage  mentioned with respect to each URL in Figure \ref{top-cat}. For example, most of the sources are used in less than 1\% of Community Notes. We additionally noticed that the category of \textit{News} has a substantial portion (almost $50\%$) of the URLs in this distribution. 


To understand the sources that dominate each category, we summarize the three most frequently cited sources across each \textbf{Type} category. Our observations indicate that most of categories depend on a few dominating sources as seen in Table \ref{tab:top_sources}, while only a few categories have a high variance in the sources as is the case with \textit{News}. 
Wikipedia's dominance within the \textit{Dictionary/Encyclopedia} category, accounting for 87.78\% of the citations in this type, highlights 
its critical role as a primary reference source. Additionally, the \textit{Fact-Checking}
category shows a substantial concentration among the top three sources (Snopes, Politifact, and FactCheck), contributing to a significant portion of the category's citations with 31.72\%, 21.42\%, and 12.13\% respectively. 

Irrespective of the category segregation, we additionally observe in Figure \ref{top-cat} that Twitter is a highly cited page, being used in notes $60168$ times ($9.3\%$ in the whole dataset). The reason is that users primarily do intra-domain fact-checking, i.e., cite other tweets in their fact-checking notes. We highlight that some of these cited tweets might have cited other sources themselves, but as we do not have access to the tweets cited or the tweets for which the community notes were written, we had to exclude Twitter tweets from our analysis. However, this does not have a significant impact on our analysis or conclusions since any links hidden in these cited tweets most likely follow the same distribution of source types that we got from our annotation stage. We also keep Twitter and X ungrouped as sources due to the ongoing discussion regarding whether the platform has changed its political leaning after Elon Musk took the company over in 2022 \cite{Bump2023, Economist2023}. The second-most cited source (after Twitter) is Wikipedia, a web encyclopedia, which in itself is a community-reviewed platform. We additionally notice that \textit{Government/Civil} and \textit{Research} sources in the top $50$ are primarily about health and nature, which might indicate that fact-checking these topics requires more expert knowledge.

\begin{figure*}[t]
\centering
\includegraphics[width=1\textwidth]{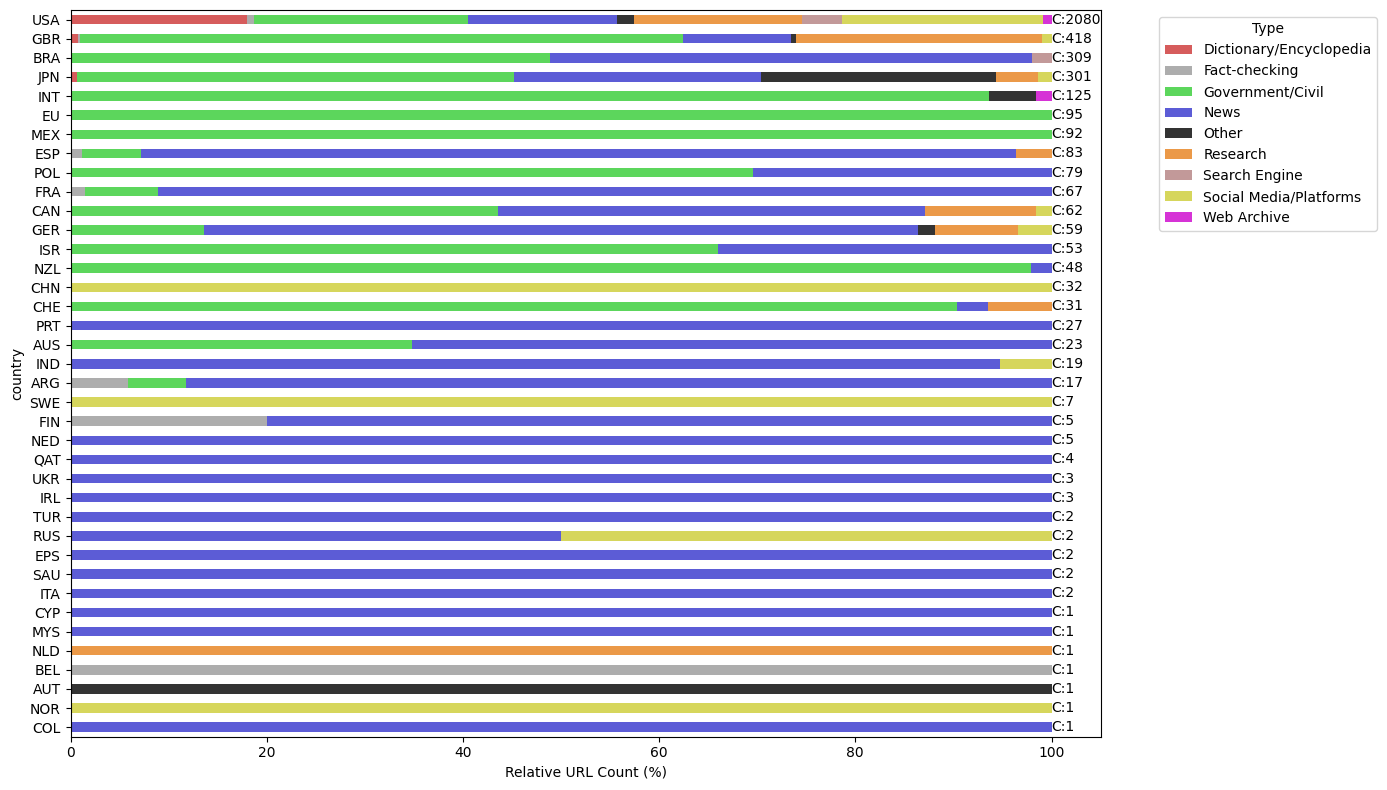}
\caption{Source Type Relative URL Count by Country. The total URL count is marked at the end of each bar. INT refers to global international sources (or sources that did not fit under any single country) and EU refers to European Union sources.}
\label{country_types}
\end{figure*}

We also categorized the top 500 sources by country which is shown in Figure \ref{country_types}. Most of the sources are from English-speaking countries, with Japan and Brazil being the most frequent from non-English countries. This is expected as Community Notes opened contributor access to users from these countries first as is highlighted in Table \ref{tab:release_dates}. 

\begin{table}[t]
\centering
\fontsize{8pt}{10pt}\selectfont
\begin{tabular}{ll}
\toprule
\textbf{Country/Region} & \textbf{Date} \\
\midrule
US & Jan 23 2021 \\
\midrule
Canada & Dec 15 2022 \\
\midrule
UK, Ireland, Australia, New Zealand & Jan 20 2023 \\
\midrule
Brazil & Mar 3 2023 \\
\midrule
Japan & Mar 21 2023 \\
\midrule
Mexico, Spain, Portugal & Apr 7 2023 \\
\midrule
Argentina, Chile, Colombia & May 4 2023 \\
\midrule
Ecuador, Guatemala, Peru, Venezuela & \\
Italy, Germany, Austria & Jun 14 2023 \\
\midrule
France, Luxembourg, Belgium, & \\
Netherlands, Switzerland, Slovakia & Jul 20 2023 \\
\midrule
Bulgaria, Croatia, Cyprus & Jul 26 2023 \\
\midrule
Czechia, Denmark, Estonia, Finland, & \\
Greece, Hungary, Iceland, Latvia, & \\
Lithuania, Malta, Norway, Poland, & \\
Romania, Slovenia, Sweden, Indonesia, & \\
Malaysia, Philippines & Nov 16 2023 \\
\midrule
Singapore, Thailand, Papua New Guinea, & \\
Brunei, Algeria, Bahrain, Egypt, Israel & Nov 22 2023 \\
\midrule
Jordan, Kuwait, Lebanon, Morocco, & \\
Oman, Palestinian Territories, Qatar, & \\
Tunisia, United Arab Emirates & \\
Hong Kong, South Korea, Taiwan & Dec 7 2023 \\
\bottomrule
\end{tabular}
\caption{Community Notes Release Dates by Country}
\label{tab:release_dates}
\end{table}


\subsubsection{Bias and factuality of sources:} To better comprehend the influence of cited sources on the political leaning and factuality of fact-checking on Twitter, we analyze the \textbf{Bias} and \textbf{Factuality} labels of these sources. In this section, we explain the annotation process of \textbf{Bias} and \textbf{Factuality} 
and give a visual overview of the \textbf{Bias} and \textbf{Factuality} distribution of the top sources (including country-wise distributions). We first annotate all of the $500$ sources with metrics for \textbf{Bias}. We do this by aggregating \textbf{Bias} labels for these sources from three media monitoring websites: \url{mediabiasfactcheck.com}, \url{allsides.com}, \url{adfontes.com}. These websites have also been widely used for the same purpose in several existing research works \cite{Rao2021,Samory_Kesiz_Abnousi_Mitra_2020,Weld2021PoliticalBA,bayiz2024susceptibility,Xiao2023}. The media monitoring platforms state that the \textbf{Bias} class is given based on the content of the pages, guest lists, and political leaning on certain topics. We aggregate these \textbf{Bias} classes using majority voting, meaning we took the dominant class over all three. If the three classes do not agree and thus no dominant class was found then we removed the source from our dataset. The labels from \url{mediabiasfactcheck.com} also had a \textit{Pro-Science} label for \textbf{Bias}, which we considered neutral as is expected from scientific sources. Additionally, as only \url{mediabiasfactcheck.com} have classes for Extreme Left and Extreme Right, we consider those classes as Left and Right correspondingly to maintain consistency across all the media monitoring websites. In the end we are left with $5$ \textbf{Bias} classes - \textit{Left}, \textit{Left-Center}, \textit{Center}, \textit{Right-Center} and \textit{Right}. The \textbf{Bias} labels of each media monitoring page and our finalized aggregated labels can be seen in Table \ref{tab:biasclasses}. After \textbf{Bias} label annotation our dataset comprises of $183$ sources which covers $991$ URLs and used in community notes $206466$ times.

\begin{figure*}[t]
\centering
\includegraphics[width=0.95\textwidth]{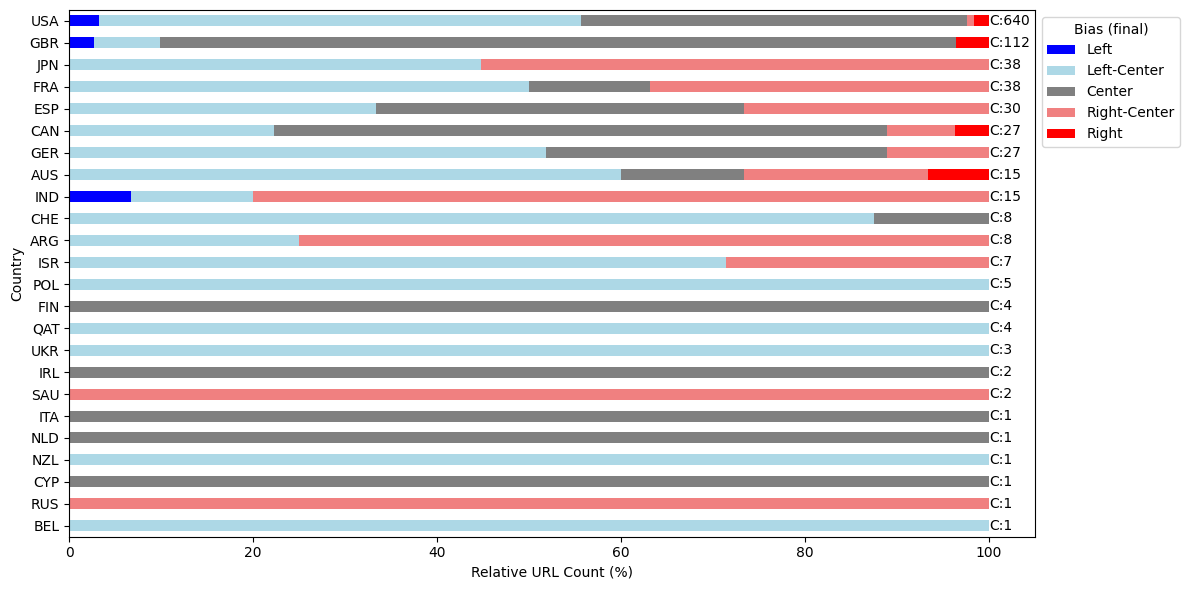}
\caption{Source \textbf{Bias} Relative URL Count by Country. The total URL count is marked at the end of each bar.}
\label{country_bias}
\end{figure*}

\begin{figure*}[t]
\centering
\includegraphics[width=0.82\textwidth]{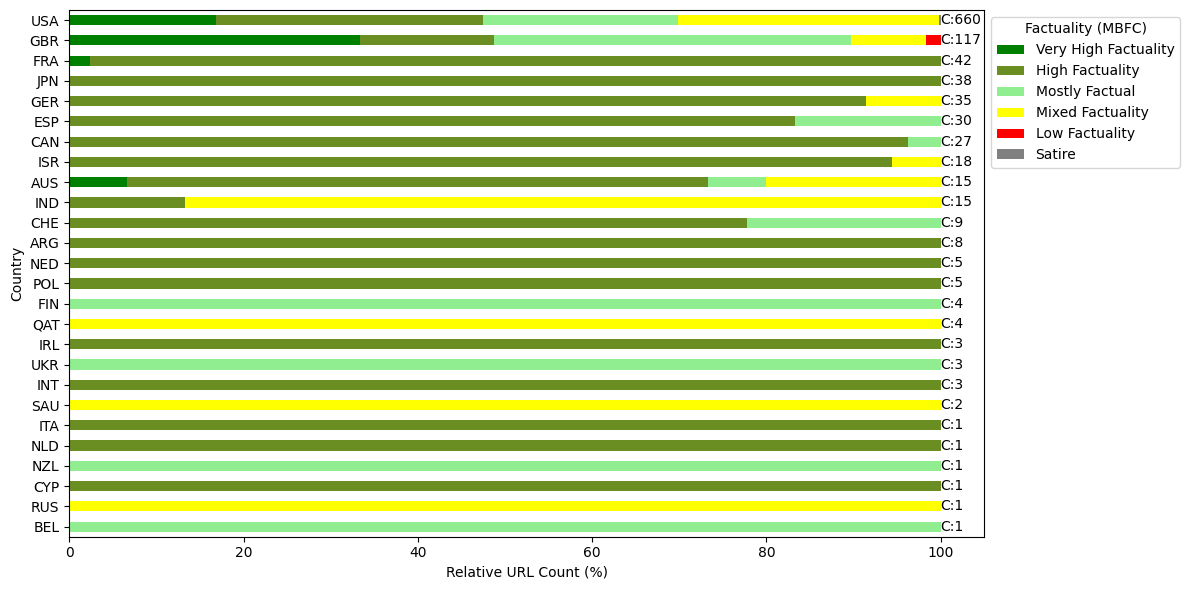}
\caption{Source \textbf{Factuality} Relative URL Count by Country. The total URL count is marked at the end of each bar.}
\label{country_fact}
\end{figure*}

\begin{table}[t]
\centering
\fontsize{8pt}{10pt}\selectfont
\begin{tabular}{llll}
\toprule
\textbf{MBFC} & \textbf{AS} & \textbf{AF} & \textbf{Bias final} \\
\midrule
Left, Extreme Left & Left & Strong Left & Left \\
Left-Center &  Lean Left & Skews Left & Left-Center \\
Center, Pro-Science & Center & Middle & Center \\
Right-Center & Lean Right & Skews Right & Right-Center \\
Right, Extreme Right & Right & Strong Right & Right \\
\bottomrule
\end{tabular}
\caption{Bias classification across media monitoring platforms (MBFC: \url{mediabiasfactcheck.com}, AS: \url{allsides.com}, AF: \url{adfontesmedia.com}). Each row aligns similar classes to a consolidated final bias label.}
\label{tab:biasclasses}
\end{table}

\begin{table*}[t]
\centering
\fontsize{8pt}{10pt}\selectfont
\begin{tabular}{lll|lll|lll}
\toprule
\multicolumn{3}{c}{\textbf{Type}} & \multicolumn{3}{c}{\textbf{Bias}} & \multicolumn{3}{c}{\textbf{Factuality}} \\
\midrule
\textbf{Type} & \textbf{Count} & \textbf{Percentage} & \textbf{Bias} & \textbf{Count} & \textbf{Percentage} & \textbf{Factuality} & \textbf{Count} & \textbf{Percentage} \\
\midrule
\textit{News} & $543$ & $54.8\%$ & \textit{Left-Center} & $448$ & $45.3\%$ & \textit{High Factuality} & $399$ & $40.3\%$ \\
\textit{Research} & $177$ & $17.9\%$ & \textit{Center} & $422$ & $42.6\%$ & \textit{Mixed Factuality} & $227$ & $22.9\%$ \\
\textit{Social Media/Platforms} & $90$ & $9.1\%$ & \textit{Right-Center} & $79$ & $8.0\%$ & \textit{Mostly Factual} & $211$ & $21.3\%$ \\
\textit{Dictionary/Encyclopedia} & $88$ & $8.9\%$ & \textit{Left} & $25$ & $2.5\%$ & \textit{Very High Factuality} & $150$ & $15.2\%$ \\
\textit{Search Engine} & $67$ & $6.7\%$ & \textit{Right} & $16$ & $1.6\%$ & \textit{Satire} & $1$ & $0.1\%$ \\
\textit{Fact-checking} & $12$ & $1.2\%$ & & & & \textit{Low Factuality} & $1$ & $0.1\%$ \\
\textit{Government/Civil} & $12$ & $1.2\%$ & & & & & & \\
\textit{Other} & $1$ & $0.1\%$ & & & & & & \\
\bottomrule
\end{tabular}
\caption{URL counts by \textbf{Type}, \textbf{Bias}, and \textbf{Factuality}}
\label{countstable}
\end{table*}

We study the distribution of the \textbf{Bias} labels by domain country origin as shown in Figure \ref{country_bias}. It's interesting to note that more polarized (Left and Right) sources were those from the USA, Great Britain, Canada, Australia and India. This is most likely because the media monitoring companies are US-based and thus also more critically evaluate English-speaking sources. Additionally, we highlight that USA-based sources are most frequent, covering 640 URLs after \textbf{Bias} annotation. 

We annotate our sources additionally with \textbf{Factuality} labels. However, as, \url{mediabiasfactcheck.com}, \url{allsides.com} and \url{adfontes.com} sources do not use the same features to identify \textbf{Factuality}, we could not aggregate these three sources and consider only \url{mediabiasfactcheck.com}. We chose \url{mediabiasfactcheck.com} out of the three as it provided maximum coverage for the sources. For \url{mediabiasfactcheck.com}, the \textbf{Factuality} label is based on the frequency of fact-checks they have passed during the last five years. There are $6$ \textbf{Factuality} classes - \textit{Very High Factuality}, \textit{High Factuality}, \textit{Mostly Factual}, \textit{Mixed Factuality}, \textit{Low Factuality} and \textit{Satire}.
However, \url{mediabiasfactcheck.com} does not provide a \textbf{Factuality} rating for all sources. Therefore, to maintain consistency in our dataset, we exclude those data points for which we do not have any \textbf{Factuality} label. After adding \textbf{Factuality} labels, our final dataset comprises of $182$ sources, covering $990$ URLs and used in $206007$ community notes. We additionally analyse the \textbf{Factuality} class across country origin of the sources as can be seen in Figure \ref{country_fact}. We can see that \textit{Low Factuality} sources are entirely from Great Britain, which also has the largest proportion of \textit{Very High Factuality}. This trend of having varied sources in terms of \textbf{Factuality} is also seen in other English-speaking countries such as the USA and Australia. 

\begin{figure*}[t]
\centering
\includegraphics[width=0.95\textwidth]{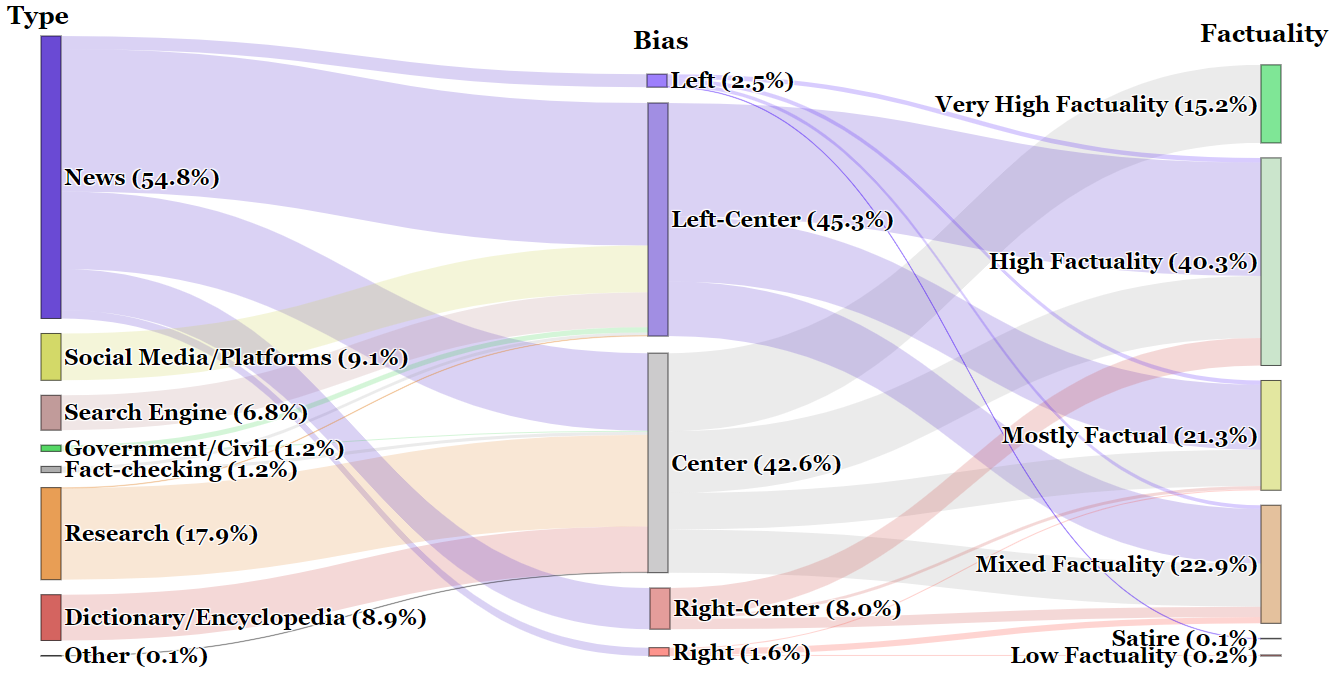}
\caption{Connectedness of source categories of \textbf{Type}, \textbf{Bias} and \textbf{Factuality}. The nodes are weighed by URL counts.}
\label{cat-sankey}
\end{figure*}

The sources after adding \textbf{Type}, \textbf{Bias} and \textbf{Factuality} labels is considered our final sources dataset, which we use to analyse the sources used in Community Notes. Every source has one \textbf{Type}, \textbf{Bias} and \textbf{Factuality} annotated category and we show their URL count distribution between categories in Table \ref{countstable}. We analyze the interrelationship between categories to understand the patterns of information framing and its impact on public perception. For example, our observations as shown in Figure \ref{cat-sankey} show the connectedness of \textbf{Type}, \textbf{Bias} and \textbf{Factuality} categories. We highlight, that the majority of \textit{News} outlets ($50.5\%$) lean towards a \textit{Left-Center} bias, and of those, a substantial amount ($79.6\%$) are highly factual.

\subsubsection{Correlation analysis:} To get a better understanding of how different categories intersect and interact with each other, we analyse the correlation of the \textbf{Type}, \textbf{Bias}, and \textbf{Factuality} categories. 
We mark the Pearson correlation coefficients as $r$ and consider scores with $r < 0.3$ weak, $0.3 < r < 0.7$ moderate and $r > 0.7$ strong. We exclude scores where $r < 0.3$ (weak correlation) from our analysis and display our results in Table \ref{tab:corr_table}.

\begin{table}[t]
\centering
\fontsize{8pt}{10pt}\selectfont
\begin{tabular}{lllll}
\toprule
& \textit{News} & \textit{Research} & \textit{Center} & \textit{Right} \\
\midrule
\textit{Center} & $-0.38$ & $0.37$ & - & - \\
\textit{Very High Factuality} & $-0.5$ & $0.5$ & $0.46$ & - \\
\textit{Mixed Factuality} & - & - & - & $0.31$ \\
\textit{Low Factuality} & - & - & - & $0.35$ \\
\bottomrule
\end{tabular}
\caption{Correlation of \textbf{Type}, \textbf{Bias} and \textbf{Factuality} categories. Shown are scores, with absolute values of r > 0.3.}
\label{tab:corr_table}
\end{table}

Our results reveal a few moderate correlations ($r > 0.3$) that hold critical implications. Firstly, \textit{Research} sources  are often categorized as both \textit{Center} ($r: 0.37$) and having \textit{Very High Factuality} ($r: 0.5$). This suggests that such sources are both reliable and seen as advocating scientific perspectives in an unbiased way. Secondly, \textit{News} sources are generally less likely to be categorized as \textit{Center} ($r: -0.38$) or possess \textit{Very High Factuality} ($r: -0.5$), which implies a potential limitation in these commonly-accessed information outlets. Thirdly, \textit{Center} sources tend to also score very high in factuality ($r: 0.46$), reinforcing the credibility of unbiased perspectives. Lastly, \textit{Right} biased sources frequently exhibit \textit{Low Factuality} ($r: 0.35$) and \textit{Mixed Factuality} ($r: 0.31$), raising questions about the credibility of such sources and their role in public discourse. 

\subsubsection{Summary of insights for RQ1:} We emphasize key points about the sources cited in notes. First, Twitter and Wikipedia are the most cited, suggesting that intra-domain fact-checking and community-reviewed content are significant in public discourse. This shows that social media platforms are not just arenas for discussion but also crucial sources of information and fact-checking. The fact-checking source \textbf{Type} category showcases a significant reliance on a few key websites, with Snopes, Politifact, and FactCheck which forms a substantial portion (65.27\%) 
of citations. The Government/Civil category, with top sources being MHLW, WHO, and Gov.uk, reflects the diversity and international representation of credible government and civil sources utilized for fact-checking. Secondly, the \textbf{Bias} and \textbf{Factuality} show most sources fall within \textit{Left-Center} (54.8\%) and \textit{High Factuality} (40.3\%), indicating a factual left-leaning fact-checking community. Thirdly, sources from English-speaking countries tend to be more frequently used and also more polarized in regards to bias and factuality. Lastly, \textit{News} sources correlate with not being \textit{Center}, \textit{Center} and \textit{Research} sources correlate positively with having \textit{Very High Factuality} and \textit{Right} biased sources with \textit{Low Factuality} and \textit{Mixed Factuality}. 
\subsection{Perceptions of audience (RQ2)}

\begin{figure*}[t]
\centering
\includegraphics[width=1\textwidth]{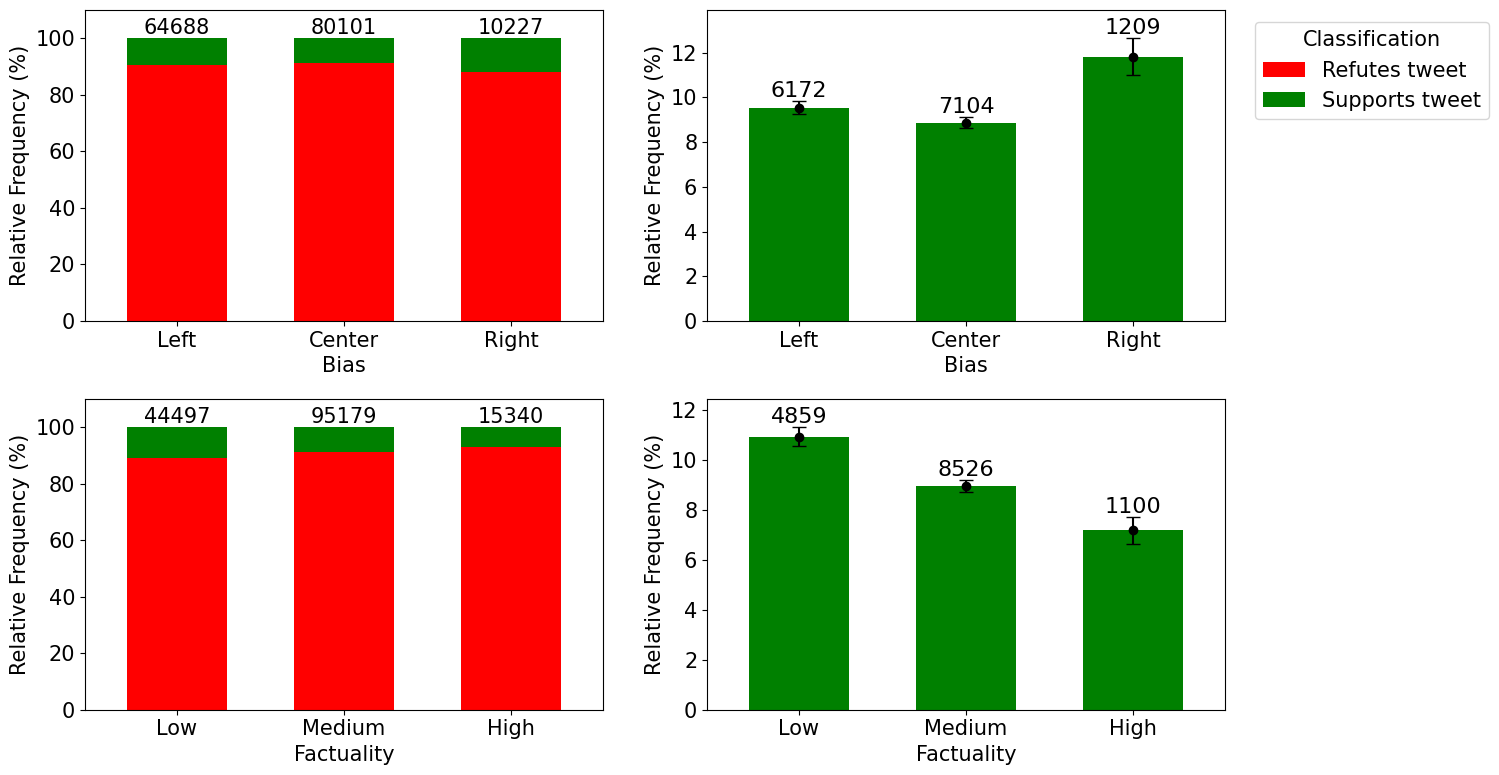}
\caption{Distribution of community notes by \textbf{Bias} and \textbf{Factuality}, categorized as either supporting or refuting Tweets. Amount of notes representing each bar is added on top. For the second plot we have also added z-score 99\% confidence intervals.}
\label{misinfo}
\end{figure*}

In this section we investigate how the \textbf{Factuality} and political \textbf{Bias} of cited sources influence the ratings and acceptance of a community note. Specifically, we look at which sources are used to support and refute notes, what is their helpfulness and how agreement levels differ in source usage. We also highlight how well the Community Note rating algorithm handles poor quality and biased content. 


As one community note can have several sources associated we need a way to aggregate \textbf{Bias} and \textbf{Factuality} labels of sources used. For this we calculate \textbf{Bias} and \textbf{Factuality} scores for each community note. The associated scores that correspond to each level can be seen in Table \ref{table:scores}. We consider \textit{Satire} equal to \textit{Very Low Factuality} score-wise as for fact-checking hidden humour can be misleading and more harmful than useful. To deal with multiple sources in community notes we disregard notes that have sources from opposite \textbf{Bias} sides and average the \textbf{Bias} scores otherwise. For \textbf{Factuality}, we average the \textbf{Factuality} scores.

\begin{table}[t]
\centering
\fontsize{8pt}{10pt}\selectfont
\begin{tabular}{ll|ll}
\toprule
\multicolumn{2}{c}{\textbf{Factuality to Score}} & \multicolumn{2}{c}{\textbf{Bias to Score}} \\
\midrule
\textbf{Factuality Level} & \textbf{Score} & \textbf{Bias Category} & \textbf{Score} \\
\midrule
Very High Factuality & 5 & Left & 2 \\
High Factuality & 4 & Left-Center & 1 \\
Mostly Factual & 3 & Center & 0 \\
Mixed Factuality & 2 & Right-Center & -1 \\
Low Factuality & 1 & Right & -2 \\
Very Low Factuality & 0 & & \\
Satire & 0 & & \\
\bottomrule
\end{tabular}
\caption{Factuality and Bias levels converted to scores.}
\label{table:scores}
\end{table}


We use a simplified 3 class system for both \textbf{Factuality} and \textbf{Bias}. The score to label transformation system is highlighted in Table \ref{table:scoretoclass}. We highlight that the largest \textbf{Factuality} class is Medium (61.4\% of notes) and the largest \textbf{Bias} class is Center (51.67\% of notes). 


\begin{table}[t]
\centering
\fontsize{8pt}{10pt}\selectfont
\begin{tabular}{llll}
\toprule
\textbf{Category} & \textbf{Score Range} & \textbf{Count} & \textbf{Percentage} \\
\midrule
Right & $< -0.5$ & 10,227 & 6.60\% \\
Center & $-0.5$ to $0.5$ & 80,101 & 51.67\% \\
Left & $> 0.5$ & 64,688 & 41.73\% \\
\midrule
Low & $< 3$ & 44,497 & 28.70\% \\
Medium & $3$ to $4$ & 95,179 & 61.40\% \\
High & $> 4$ & 15,340 & 9.90\% \\
\bottomrule
\end{tabular}
\caption{Distribution of Bias and Factuality scores with labels, counts, and percentages.}
\label{table:scoretoclass}
\end{table}

\begin{figure*}[t]
\centering
\includegraphics[width=1\textwidth]{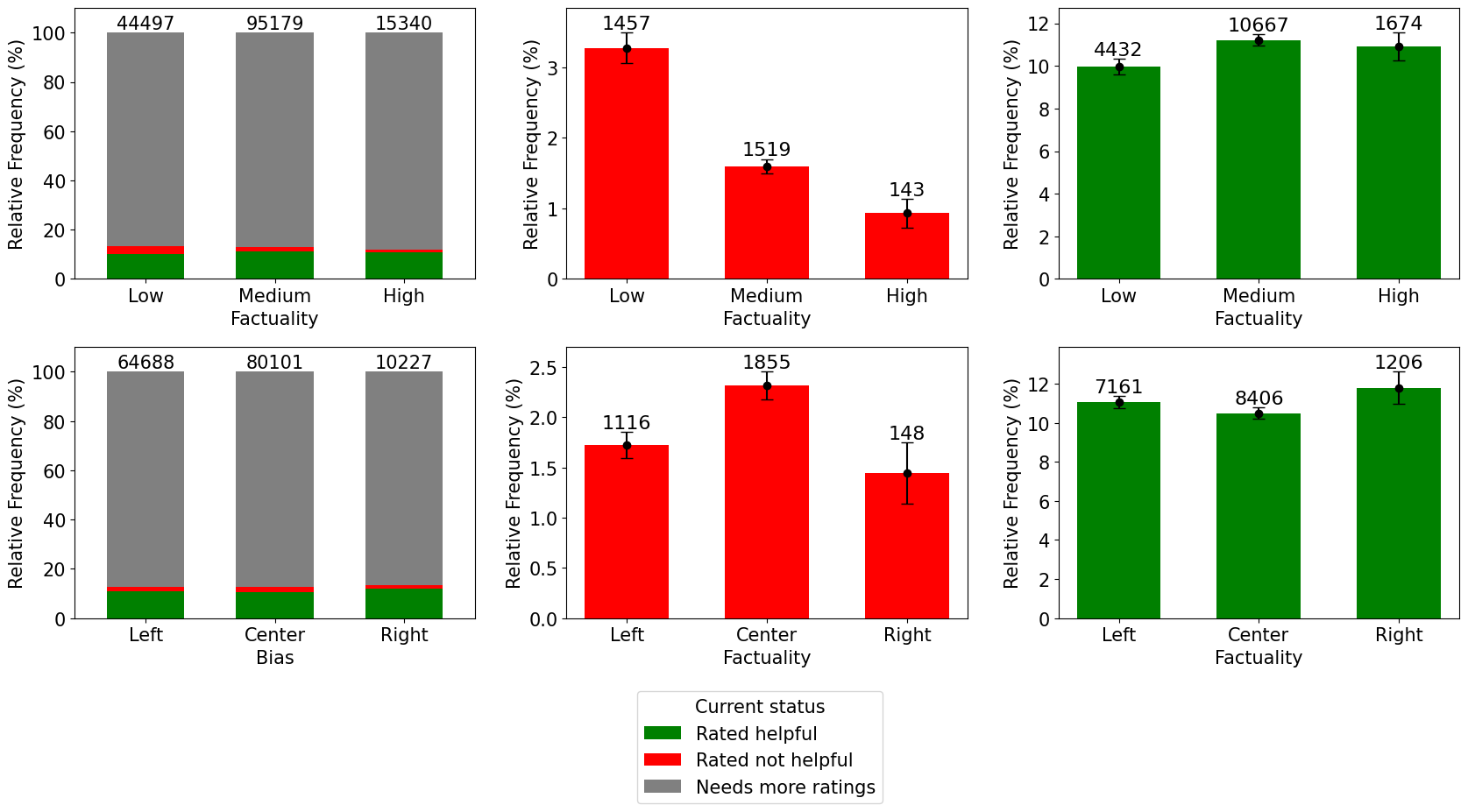}
\caption{Distribution of community notes by \textbf{Bias} and \textbf{Factuality}, categorized as either helpful, not helpful or needs more ratings. Amount of notes representing each bar is added on top. For the second and third plot we have also added z-score 99\% confidence intervals.}
\label{helpful}
\end{figure*}

\subsubsection{Role of sources in supporting or refuting the content of tweets:} Community notes can be used to \textbf{support} both the truthfulness of a tweet (marking the original tweet as not misleading) or \textbf{refute} it (marking the original tweet as misleading). This label is given by the community note writer. We aim to understand which categories of sources are used more for supporting and which ones are for refuting tweets. We can see the 
\textbf{Bias} and \textbf{Factuality} distributions of the refuting/supporting notes on Figure \ref{misinfo}.

On \textbf{Bias}, we can see that notes with Right-wing sources are used relatively more when supporting tweets compared to notes that use Center or Left-wing sources ($p < 0.01$). This could indicate a broader tendency within the conservative media space to create self-reinforcing loops of information \cite{Jay2020}. 

Our observations on \textbf{Factuality} indicate that sources with lower factuality are used relatively more to support tweets than those with high factuality ($p < 0.01$). This can be an indication of misinformation enforcement, 
where a community note has been written to a misleading note to make it seem credible.

\subsubsection{Role of sources in community note helpfulness:}
Community Notes undergo a contributor-driven rating process to determine their status as "helpful", 
not helpful", or "needs more ratings", affecting their visibility on site timelines and posts \cite{NoteRanking2024}. Initially, all notes start in a "Needs More Ratings" state until receiving at least five ratings, at which point they may be classified as helpful or not helpful. Notes identified as fact-checking potentially misleading tweets that meet specific helpfulness score criteria are marked as helpful and displayed on posts, whereas those not meeting the criteria are deemed not helpful. The process includes a diligence scoring mechanism to evaluate the accuracy and sourcing of information, ensuring that notes recognized as reliable and clear by a broad spectrum of users are highlighted. We can see the \textbf{Bias} and \textbf{Factuality} distributions of the helpful and not helpful notes in Figure \ref{helpful}.

We aim to look at the \textbf{Bias} and \textbf{Factuality} of sources used in helpful and not helpful notes. When it comes to \textbf{Bias}, we notice that Left and Right sources are associated relatively more frequently  with helpful than not helpful notes compared to Center sources ($p < 0.01$). This might be due to people having a confirmation bias when looking at content confirming their views.

However, regarding \textbf{Factuality} we notice a trend where notes that hold low factuality sources are generally rated less helpful than those with medium or high factuality sources ($p < 0.01$). This confirms that notes with lower quality sources are effectively classified by the Community Notes Ranking algorithm. The \textbf{Factuality} distribution of the helpful and not helpful 
notes is shown in Figure \ref{helpful}.


\subsubsection{Perceived agreement by source categories:} We analyse the perceived agreement levels of the Community notes per source category. For this, we use the existing ratings associated with each note and created an agreement index using the number of ratings that agreed and disagreed with the note:

\begin{figure*}[t]
\centering
\includegraphics[width=1\textwidth]{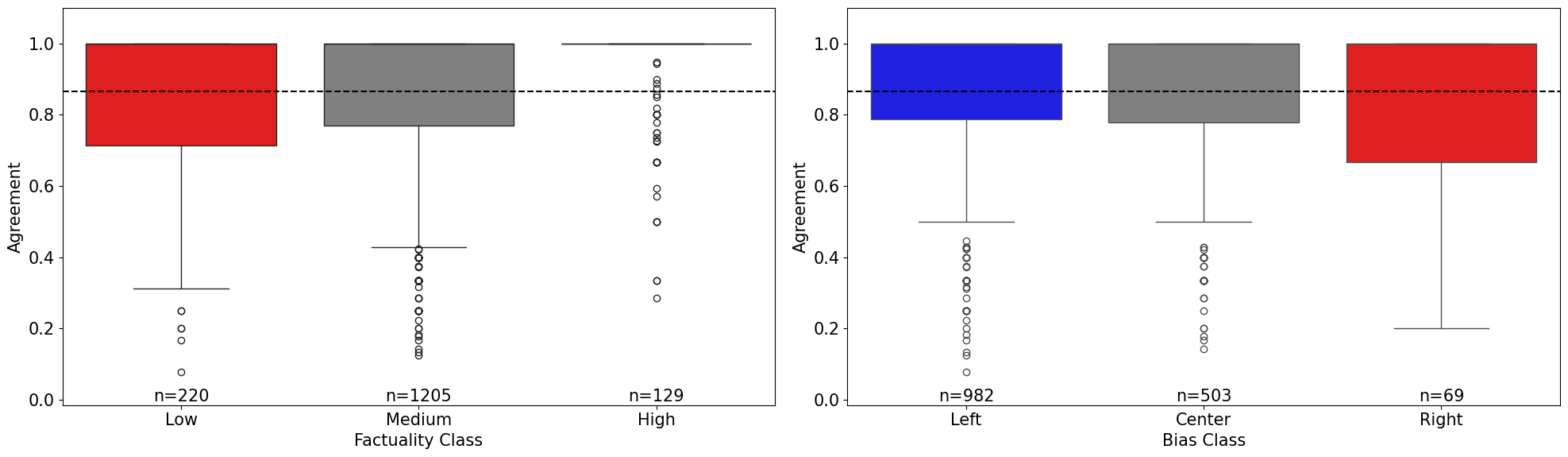}
\caption{Distribution spread of aggregated agreement by factuality and bias categories. The black dotted line indicates average agreement.}
\label{agree}
\end{figure*}

\begin{equation}
\text{{Agreement}} = \frac{{\text{{Agree}}}}{{\text{{Agree}} + \text{{Disagree}}}}
\end{equation}

We indicate a threshold of $0.5$, where scores greater than $0.5$ indicate that the notes are more agreeable than disagreeable, while scores lower than $0.5$ suggest the opposite. We plot the distribution spread of aggregated agreement for our categories of \textbf{Bias} and \textbf{Factuality} in Figure \ref{agree}. We also notice that the average agreement per our community notes is 0.87.


We explore whether the bias of the cited sources in notes correlates with the level of agreement those notes receive. We assume that notes citing more politically neutral sources will attract broader agreement. This is based on the assumption 
that neutral sources may be less likely to polarize opinion compared to clearly left- or right-leaning sources \cite{mitchell2014polarization}. On analyzing the \textbf{Bias} category results, our initial observations indicate that notes that cite Right sources have significantly lower agreement levels, with the lower quartile showing an agreement score under $0.5$. However, notes that cite Left sources show similarly high levels of agreement to the notes that use Center sources.
These findings highlight that notes which cite more politically center or left leaning sources are generally more agreeable than those relying on Right sources.


Factuality plays a large role in the agreeableness of statements; thus, we expect that sources with higher \textbf{Factuality} ratings are also generally more agreeable. When we inspect the \textbf{Factuality} category agreement levels, we see that higher \textbf{Factuality} sources, also have generally higher aggreement levels and vice versa for lower \textbf{Factuality} sources. Notes that cite sources of Low factuality see the lowest agreement levels, with their lower quartiles being below the 0.5 level. This indicates that community note users pay attention to the factuality of the source when rating the notes.

\subsubsection{Summary of insights for RQ2:} We summarize the main takeaways of this analysis next. Our results shows a significant preference for right-leaning sources to support the content of tweets which hints at a tendency among conservative media outlets to form self-reinforcing information cycles. Similarly, we found that community notes that support tweets often rely on sources with low factuality scores, therefore highlights a systemic issue with misinformation. Further, our study indicates that notes associated with either left or right sources tend to be considered helpful, whereas those that cite sources of lower factuality are not, demonstrating the Community Note rating algorithm's ability to effectively filter content quality. Finally, we note that community notes citing more neutral or factually sound sources receive higher agreement levels, emphasizing the importance of source quality in achieving community agreement.  
\section{Conclusions}
\label{sec:conclusion}

Our investigation into Twitter's Community Notes has uncovered distinct patterns in the use of sources for community-led fact-checking, showing clear trends and biases. We've discovered that Twitter and Wikipedia are often the go-to sources, highlighting a preference for checking facts within the platform and relying on community-reviewed information for public discussions. Our findings reveal that sources are mainly left-center in bias and high in factuality, suggesting a left-leaning trend in the fact-checking community. Moreover, sources from English-speaking countries are used more frequently, indicating a bias towards these regions and a more pronounced polarization in terms of political bias and factual accuracy. This polarization is especially evident in the types of sources cited, with news sources often showing clear bias, whereas academic and research sources are typically linked to very high factual content. In contrast, right-biased sources are often associated with lower levels of factuality. Adding to this, our results indicate a noticeable preference for right-leaning sources to support tweets, which may suggest that right-wing users are creating echo chambers on the platform. We also observed that notes endorsing tweets often depend on less factual sources, pointing to a broader issue with misinformation. Interestingly, our analysis shows that notes linked to both left and right biases are usually seen as helpful, except when they reference lower-quality sources. The low agreement of low-quality sources justifies the usage of ratings of notes used by the Community Note rating algorithm in sifting through content quality. Additionally, notes that cite more balanced or factually accurate sources tend to receive higher levels of agreement, underscoring the critical role of source quality in fostering community consensus.


\section*{Ethics statement}
\label{sec:ethics}

We emphasize that we rely entirely on publicly available and anonymous data shared on Twitter Community Notes; hence, we do not and are unable to obtain consent from users who wrote notes to tweets. We add that this research did not involve any human subjects or crowd-workers, and all annotations were done by the authors of the work itself using reliable materials. We discuss the data handling in depth to mitigate any misuse of our results.

\begin{acks}
This work has received funding from the EU H2020 program under the SoBigData++ project (grant agreement No. 871042), by the CHIST-ERA grant No. CHIST-ERA-19-XAI-010, (ETAg grant No. SLTAT21096), and partially funded by HAMISON project.
\end{acks}

\bibliographystyle{ACM-Reference-Format}
\bibliography{main.bib}

\end{document}